\documentclass[aps,prb,12pt,eqsecnum,showpacs]{revtex4}
\usepackage{graphicx}
\usepackage{amssymb}
\usepackage[usenames]{color}
\begin{document}
\title{Heterogeneous diffusion, viscosity and the Stokes Einstein relation in binary liquids}
\author{H. R. Schober}
\affiliation{Peter Gr\"unberg Institut, Forschungszentrum J\"ulich, 52425 J\"ulich, Germany}
\author{H. L. Peng}
\affiliation{Institut f\"ur Materialphysik im Weltraum, Deutsches Zentrum f\"ur Luft- und
Raumfahrt (DLR), 51170 K\"oln, Germany}
\begin{abstract}
We investigate the origin of the breakdown of the Stokes-Einstein relation (SER) between
diffusivity and viscosity in undercooled melts. A binary Lennard-Jones system, as a
model for a metallic melt, is studied by molecular dynamics. A weak
breakdown at high
temperatures can be understood from the collectivization of motion, seen in the
isotope effect. The strong breakdown at lower temperatures is connected to an increase
in dynamic heterogeneity. On relevant timescales some particles diffuse much faster than
the average or than predicted by the SER. The van-Hove self correlation function 
allows to unambiguously identify
slow particles. Their diffusivity is even less than predicted by the SER. The time-span
of these particles being slow particles, before their first conversion to be a fast one, is larger than the decay time of the stress correlation.
The contribution of the slow particles to the viscosity rises rapidly upon cooling.
Not only the diffusion but also the viscosity shows a dynamically heterogeneous 
scenario. We can define a ``slow'' viscosity. The SER is recovered as relation
between slow diffusivity and slow viscosity. 
\end{abstract}
\date{\today}
\baselineskip3.5ex
\maketitle
\
\section{Introduction}
Diffusivity and shear viscosity largely characterize the dynamics of liquids. The 
shear viscosity $\eta$ is a macroscopic measure of the resistance of the fluid against
shear deformation whereas the diffusion coefficient $D$ measures the long range atomic
motion. Far above the liquidus temperature the atomic quantity $D$ and the macroscopic
$\eta$ are connected by the Stokes-Einstein relation (SER):\cite{einstein}
\begin{equation}
D(T) = \frac{k_B T}{c \eta(T) \pi R},
\label{SER_eq}
\end{equation}
where $T$ is the temperature, $R$ an effective radius of the particle, $k_B$ is the
Boltzmann constant and the constant $c$ varies between 4 and 6 depending on slip or 
stick boundary between particle and fluid.
The SER is derived
for the diffusion of uncorrelated macroscopic spheres in a liquid. Treating the motions
of the solvent atoms as uncorrelated the SER is also applied to the diffusion of 
single atoms or molecules at high temperatures. Discrepancies of up to 20\% can be 
absorbed in an effective hydrodynamic radius $R_H$ and a change of the boundary condition
from stick to slip.\cite{balucani} In the absence of values of either $\eta$ or $D$
the SER is often employed to estimate the missing quantity. It has been widely used 
in fields as distinct as transport in cells\cite{clough:82,mccarthy:01} and magma 
flow \cite{poirier:88,poe:97}. Molecular transport is treated by using an effective
hydrodynamic radius in Eq.~(\ref{SER_eq}). The shape of small molecules can be included
by replacing the Stokes formula for spheres by the one for ellipsoids.\cite{perrin:34}  
Some effects of collective
motion can be included by accounting for the wave vector dependence of the velocity
field leading to a generalized SER.\cite{balucani,gaskell:89} 

The viscosity $\eta$ is not always known, or is difficult to compute and, therefore, 
the structural relaxation time $\tau$ is often used as an alternative to study the 
temperature
dependence. This alternative equation is frequently referred to as Stokes-Einstein-Debye
relation (SEDR), assuming $\eta(T) \propto \tau/T$
\begin{equation}
D(T) \tau(T) = {\rm const}. 
\label{SEDR_eq}
\end{equation}
The proportionality $\eta(T) \propto \tau/T$ holds approximately, but breaks down even
for simple binary glasses as the temperature is lowered.\cite{shi:13}

It has been shown both in experiment \cite{roessler:90,fujara:92,cicerone:95,heuberger:96,voronel:98,meyer:03,swallen:03,bartsch:10,brillo:11} and simulation 
\cite{thirumalai:93,nicodemi:98,angelani:98,yamamotu:98,allegrini:99,demichele:01,mukherjee:02,bordat:03,kumar:06,das:08,affouard:09,HS:11,HS:16,lu:12,shi:13,sengupta:13,sengupta:14,jaiswal:15} that upon cooling towards the glass transition the SER
breaks down, the diffusivity remaining much larger than estimated via the 
SER or SEDR from the increase in viscosity or relaxation time. These experiments and 
simulations report mostly a breakdown in the SER about 30\% above the glass transition 
temperature or
in the region of the critical temperature $T_c$ of mode coupling theory\cite{goetze:92}.
Both measurements of metallic glasses \cite{meyer:03,brillo:11} and computer simulations
\cite{das:08,HS:11,HS:16,sengupta:13,jaiswal:15} have however shown that there is 
already a weaker breakdown
of the SER at much higher temperatures. This has been attributed to an increase of
cooperativity.
 
To account for deviations 
from the SER, an empirical modification, fractional SER,has 
been proposed, where $\eta$ is replaced by $\eta^p$ with $p<1$.\cite{pollack:85}  
A number of theories predict fractional SER's or SEDR's. There is no consent on the
value of the fractional exponent $p$. It 
has been argued that such deviations from $p=1$ should be taken as a hint to look
for effects beyond hydrodynamics which affect $D$ or $\eta$ exponentially.\cite{zwanzig:85} One such effect could be the increased collectivity of motion. 
In that temperature range, barriers in configuration space start to dominate the
dynamics. Hopping motion becomes visible and eventually dominates over flow. 
Consequently the diffusion coefficient rises above its SER-value. 
Concentrating on
hopping motion between deep energy minima (meta-basins) Heuer and coworkers found for
a fractional SEDR exponents as low as $p \approx 0.3$.\cite{heuer:08}   
Considering fluctuations in the heights of barriers in the hopping dynamics of a
colloidal system larger values of $p \approx 0.8$ have been reported.\cite{schweitzer:04}
Similar values are obtained in a model where one treats fluctuations in the time
between hops, persistence times, again similar exponents are gained.
These values are in good agreement with experiments.\cite{fujara:92,swallen:03} 
In a simulation treating fluctuations of jamming of atomic mobility a similar exponent, 
$p=0.73$, was found.\cite{jung:04} 
Another argument for a fractional
SER exponent can be given using Adam-Gibbs theory arguing that the activation free
energies of diffusion and viscosity are differently influenced by configurational
entropy.\cite{sengupta:13} 

A fractional SER can also be derived from the coupling model\cite{ngai:99}.
 
In an experiment on Zr$_{64}$Ni$_{36}$ upon cooling a transition 
from the SER dependence $D\eta \propto T$ to $D\eta ={\rm const}$
has been observed. It was
rationalized from Mode coupling theory\cite{goetze:92} (MCT) that describes 
the transition from a flow motion
dominated by binary collisions to one dominated by collective motion.\cite{brillo:11} 
Surprisingly this transition sets in at a temperature far above the MCT
critical temperature, even above the liquidus temperature.
Including spatially heterogeneous relaxations in MCT again a fractional SER is
found.\cite{rizzo:15}

The low temperature breakdown of the SER is mostly ascribed to the dynamic
heterogeneity that abruptly grows at similar temperatures. To quantify the
effect has proved rather elusive and often somewhat arbitrary definitions of
fast and slow particles have been used.
Particles are often
divided into fluid or solid types. Solid type particles diffuse via hopping
motion which is often regarded as nearest neighbor hopping.
For a hard sphere model 
with density above the MCT critical density it was argued that
the breakdown is due to particles that hop over distances that are integers of
the particle spacing (solid like particles).\cite{kumar:06} Alternatively the
appearance of secondary peaks in the self correlation function is used.\cite{sengupta:14}
These secondary peaks are mainly observed for small minority components. We will
show that the SER breakdown also occurs for the majority component where no
secondary peak is found and which dominates the viscosity. In agreement with the
experimental results on diffusion in metallic melts\cite{RMP} no typical length scale
was found in simulations of CuZr.\cite{KS:04} Hopping motion can be identified but
does not involve a definite length scale and in general cannot easily be identified 
from the atomic self correlation function. In a metallic melt, jump processes are 
typically not jumps between localized sites but jumps of strings of atoms.\cite{RMP}

In composite liquids the breakdown of the SER occurs in general at different 
temperates for the different components.\cite{affouard:09,HS:11} It has been argued 
that the breakdown of the SER is directly related to a dynamical decoupling of the 
components.\cite{affouard:09} Such decoupling is in contradiction to the picture of
collective string motion. Changes in dynamics, however, do affect the two components
differently and quantities such as the ratio of the diffusion coefficients can be
used as marker.\cite{HS:11,jaiswal:15}

In the following we present a molecular dynamics study of a binary Lennard-Jones system 
for temperatures down to approximately the MCT critical temperature. After giving
calculational details we evaluate
the diffusion coefficients and the viscosity and find a breakdown of the SER. From
the ratio of the component diffusion coefficients we find two transition temperatures,
first weak break around $2*T_c$ and then a strong violation of the SER near $1.2*T_c$. We then establish
that diffusion in the relevant temperature interval is collective and that heterogeneity
rapidly rises at the low temperatures. The van-Hove self correlation function is used to 
identify slow particles without using ad hoc cutoffs. The evolution of the slow particle 
contribution defines a slow diffusion coefficient and the lifetime of a particle staying
slow. This time is larger than the average stress correlation time. Like the
diffusion the viscosity is subject to a dynamically heterogeneous scenario. For a virtual
fluid, formed by these slow particles only, the SER is recovered.

\section{Calculational details}
The simulations were done for binary systems of 5488 atoms with a ratio of 4:1 between
A- and B-atoms. The atoms interact via a  
binary Lennard-Jones potential described by
\begin{equation}
V_{ij}(R) = 4 \epsilon_{ij} \left[ \left( \sigma_{ij} /R \right)^{12} - 
                         \left( \sigma_{ij} /R \right)^{6} +
                          A_{ij} R + B_{ij} \right] .
\end{equation}
where the subscripts $ij$ denote the two species A and B.
The potential cutoff was set at $R^c_{ij} = 2.5\sigma_{ij}$. As the parameters, we
took the values of Kob and Andersen \cite{kob:95}:
$\epsilon_{AA}= \epsilon = \sigma_{AA} =\sigma = 1$, 
$\epsilon_{BB}=0.5$, $\sigma_{BB}=0.88$,
$\epsilon_{AB}=1.5$ and $\sigma_{AB}=0.8$.  
The parameters
$A_{ij}$ and $B_{ij}$
ensure continuity of the potential and its first derivative at the
cutoff.
All masses are set to $m_j=1$.
As usual, in the following,  we will give all results in the 
reduced
units of energy $\epsilon$, $\sigma$, and atomic mass. To compare with real 
metallic glasses one can equate one
time unit ($(\epsilon/m\sigma^2)^{-1/2}$) roughly to 1~ps. 
The time step is 
$\Delta t = 0.005$. At the higher temperatures the time step was reduced to 
$\Delta t = 0.001$ and $\Delta t = 0.00025$.
The calculations were done with periodic boundary conditions at constant volume, 
where the volume at each temperature 
was fixed to give an average pressure, after aging, of $p=5\pm 0.05$.
The heat bath is simulated 
by comparing
the temperature averaged over  20 time steps with the nominal temperature.
At each  time step 1\% of the temperature difference is adjusted by
random additions to the particle velocities. Apart from the very first
steps of the aging procedure the correction, after excursions
of the temperature due to relaxations, does not exceed $10^{-4}$ of the
average velocity. 
This procedure assures that
existing correlations between the motion of atoms are only minimally
affected. 
The investigated temperatures ranged from $T=2$ to  $T=0.47$.
The samples were aged at the high temperature then
rapidly quenched to the next lower temperature and again aged. Apart from small residual
effects at the lowest temperatures $T=0.48$ and $T=0.47$ no significant effect of not
sufficient aging was observed. 
We used 8 independent samples. To improve the statistics for some calculations, e. g. the
diffusional isotope effect and viscosity, these samples were split, after aging, into up to 250
sub-samples each. These were subsequently aged for a shorter time span.

\section{Calculation}

\subsection{Diffusion}

\begin{figure}[htb]
\includegraphics*[bb=40 100 540 549,width=10cm,clip,keepaspectratio]{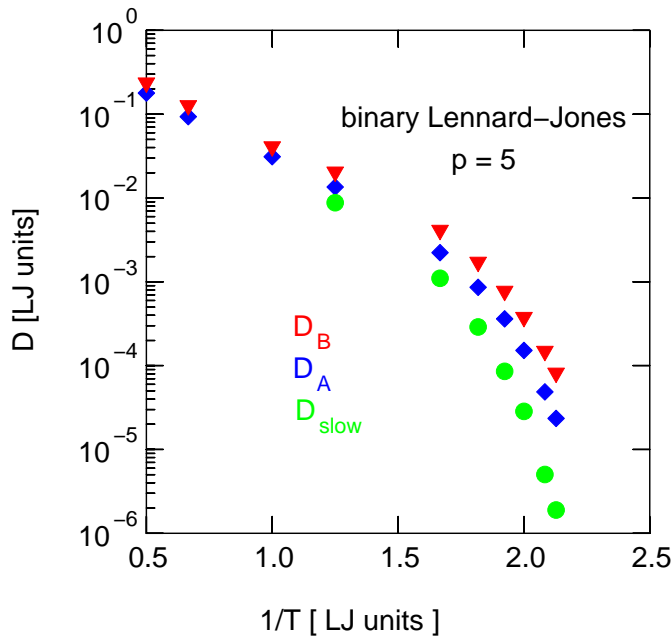}\\[0.5cm]
\caption{Diffusion coefficients versus temperature: blue diamonds: A-atoms, red 
triangles: B-atoms, green circles: slow diffusion of A-atoms.}
\label{diffusion_fig}
\end{figure}

We calculate the diffusion coefficients from the long time evolution of the mean square
displacements using the standard expression
\begin{equation}
D_\ell(T) = \lim_{t\to\infty} \langle s^2(t) \rangle_\ell /6t ,
\label{diffusion_eq}
\end{equation} 
where $ \langle\rangle_\ell$ indicates the average over all atoms of species $\ell$ and all samples.
Fig.~\ref{diffusion_fig} shows for both A- and B-atoms the usual behavior. At high
temperature the diffusivity follows with temperature an  Arrhenius  law. 
As predicted by MCT or a Vogel-Fulcher-Tamann (VFT) relation, at lower temperatures the
diffusivities drop rapidly below their respective Arrhenius values. The onset temperature for
this drop is $T_s \approx 0.6$ for both components.
The temperature range of the
present investigation does not allow an unambiguous identification of  low temperature
VFT or MCT laws. The MCT critical temperature is $T_c \approx 0.48$. Additionally, we show in  the figure values of the slow diffusion 
contribution 
($D_{slow}(T)$), which we will discuss further down.
To get an estimate for the timescale of diffusional motion we
define a diffusional time, $t_D(T)$ as the time in which the average mean square 
displacement
increases, according to Eq.~(\ref{diffusion_eq}),  by $\sigma^2$. In the temperature 
interval from $T=2$ to $T=0.47$ this diffusional time of the A-atoms increases from 
$t_D \approx 1$ by
four orders of magnitude to about 7000. The values of $t_D(T)$ are given in 
Fig.~\ref{times_fig} further down.

\begin{figure}[htb]
\includegraphics*[bb=40 100 540 549,width=10cm,clip,keepaspectratio]{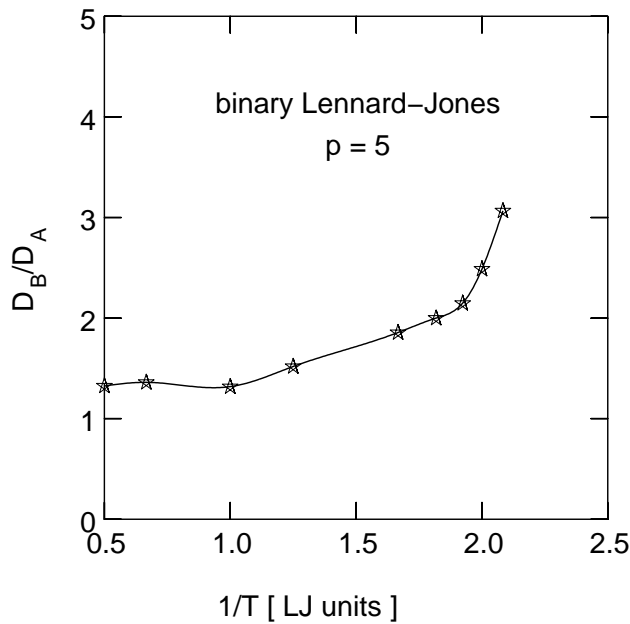}\\[0.5cm]
\caption{Ratio of the diffusion coefficients of the A and B components versus inverse temperature. The line is a guide to the eye.}
\label{dadb_fig}
\end{figure}

It has been argued \cite{HS:11,jaiswal:15} that in binary metallic melts the temperature 
dependence of the ratio of the diffusion coefficients of the two components is a 
sensitive probe of the change in dynamics which affects the two components differently. 
According to Fig.~\ref{dadb_fig} the present system shows three temperature
regimes. At high temperatures ($T>1$) the two diffusion coefficients evolve in parallel.
In an intermediate regime ($1>T>0.5$) the ratio increases with $1/T$. Finally, below 
$T=0.5$ the ratio $D_B/D_A$ increases rapidly. The high temperature regime as
expected for simple liquid when diffusion 
is dominated by binary collisions and back-flow effects or chemical bonds are not too
important.
The ratio $D_B / D_A$ is given by the inverse ratio of the effective atomic radii. 
In this temperature regime  the SER is expected to hold.

In the other two temperature regimes one expects first a 
weak deviation and then a catastrophic breakdown of the SER. This will be discussed 
further down.

\subsection{Viscosity}

We calculate the shear viscosity, $\eta(T)$, from the Green-Kubo relation as time 
integral over the stress auto-correlation function $\hat{\eta}(T,t)$ \cite{allen:87}:
\begin{equation}
\eta(T) = \int_0^\infty \hat{\eta}(T,t) dt
\label{eta_eq}
\end{equation}
with
\begin{equation}
 \hat{\eta}(T,t) = \frac{1}{k_b T V}\langle \sigma^{xy}(t)\sigma^{xy}(0)\rangle
\label{etat_eq}
\end{equation}
where $V$ is the simulation volume and $\sigma^{xy}$ stands for the off-diagonal elements
of the macroscopic stress tensor computed from the momenta and virials
\begin{equation}
\sigma^{xy} = \sum_{i=1}^N \left( m_i v_i^x v_i^y - 
                          \sum_{j>i}{
                         \frac{\partial V_{ij}}{\partial r_{ij}} 
                         \frac{r_{ij}^x r_{ij}^y} {r_{ij}}}
\right).
\label{stress_eq}
\end{equation}
\begin{figure}[htb]
\includegraphics*[bb=40 40 550 410,width=8cm,clip,keepaspectratio]{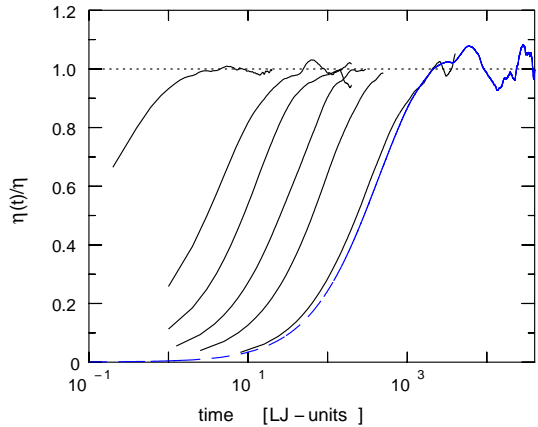}\\[0.5cm]
\caption{Normalized stress integral, Eq.~(\ref{eta_eq}) as function of integration time.
Solid lines: temperatures from left to right $T=$1, 0.6, 0.55, 0.52, 0.50, 0.48; blue
independent control samples at $T=$ 0.48.}
\label{eta_log_fig}
\end{figure}
Fig.~\ref{eta_log_fig} shows the time evolution of the Green-Kubo integral for 
temperatures from $T=1$ down to $T=0.48$. AS an additional check of 
the reliability of our results we compared them for all temperatures to 
the values obtained for sets of different samples which were
fully independent of the ones used in the present study. These 
control calculations were done using the LAMMPS program
package.\cite{lammps} The agreement was always within 5\%. We are concentrating
in this work on the strong violation of the SER at lower temperature where it
is Larger by orders of magnitude.

\begin{figure}[htb]
\includegraphics*[bb=40 40 550 550,width=8cm,clip,keepaspectratio]{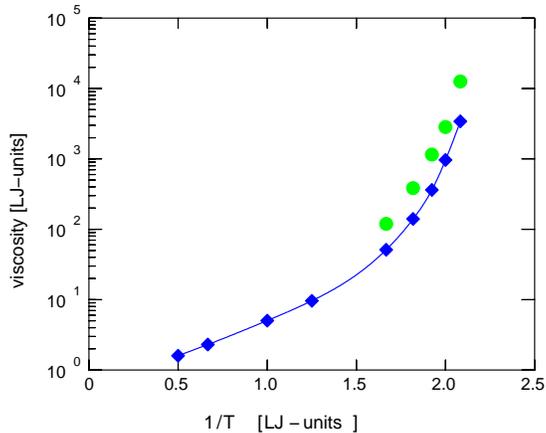}\\[0.5cm]
\caption{Viscosity as function of inverse temperature (blue diamonds). The green circles show
the viscosity of the virtual slow system. The line is a guide to the eye}
\label{viscosity_fig}
\end{figure}

The blue diamonds in Fig.~\ref{viscosity_fig} show the viscosity calculated from 
Eq.~(\ref{eta_eq}). As 
characteristic for metallic melts, two temperature regimes can be distinguished: a slow
increase upon cooling at high temperatures and a  much more rapid one at lower
temperatures. The strength of the change of the asymptotic slopes shows that our system
can be classified as fairly fragile.
By extrapolation we find a crossover temperature of $T\approx 0.57$ in
good agreement with the onset temperature found for diffusion. The transition from
high to low temperature viscosity spans the temperature interval $0.8 <1/T<0.5$. This
corresponds roughly to the intermediate regime in $D_B/D_A$, Fig.~\ref{dadb_fig}.      

\subsection{Stokes Einstein Relation}

\begin{figure}
\includegraphics*[bb=40 40 550 550,width=10cm,clip,keepaspectratio]{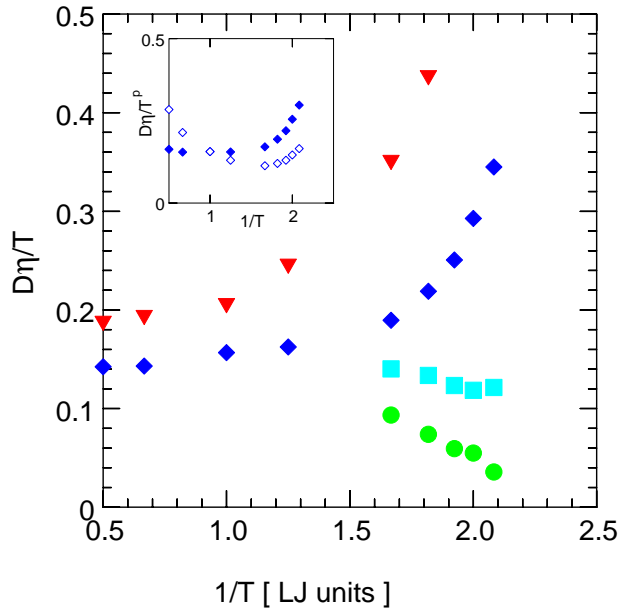}\\[0.5cm]
\vspace{-1.5cm}
\caption{Stokes Einstein relation against inverse temperature for A-atoms 
(blue diamonds), B-atoms (red triangles), slowly diffusing A-atoms (green circles) and 
a virtual slow system (cyan squares). The insert shows for the A-atoms the temperature
dependence of $D \eta / T^{p}$ for $p=0.8$ (full blue diamonds) and $p=0$ (open blue 
diamonds.} 
\label{SER_fig}
\end{figure}

Combining the diffusion and viscosity data we can now calculate the SER, 
Eq.~(\ref{SER_eq}). In Fig.~\ref{SER_fig} we plot $D\eta /T$ against inverse temperature.
The SER holds as long as $D\eta /T$ remains constant, independent of temperature.
In agreement with the constant ratio $D_B/D_A$ the SER holds for temperatures down to
about $T=1$ for both components. Below $T=0.52$ one clearly observes a rapid increase. 
The intermediate region is not seen unambiguously. For the larger A-atoms the relation
$D \eta \approx {\rm const}$ holds approximately the temperature interval $ 0.8<1/T<2$
(open blue diamonds in the insert). This agrees
with the experimental observation of Brillo et al. \cite{brillo:11} and simulations
for CuZr\cite{HS:11,HS:16}. For the smaller B-atoms this transition interval seems 
shifted 
to higher temperatures $ 1<1/T<1.7$ (not shown). It has been argued that the near 
constancy of 
$D \eta$ is a signature of collective flow as described by MCT. However, a fractional
SER, $D \eta / T^{0.8}$, (solid blue diamonds in the insert) holds in the shifted 
temperature interval $2<1/T<1.5$. This value of $p=0.8$ is in good agreement
to experiments on small organic molecules.\cite{fujara:92,swallen:03}. 
The present data are insufficient to clearly
identify the proper relation in this higher temperature range $T>0.6$ ($1/T<1.6$).

Also shown in the figures are values for slow atoms. This will be explained in the
following section where we discuss the evolution of diffusion and viscosity with
temperature in more detail.

\section{Discussion}

To gain more insight into the breakdown of the SER we take a closer look at
both diffusion and viscosity. Main reasons for the breakdown, discussed in the 
literature, are increases of both collectivity and dynamic heterogeneity upon cooling.
To quantify the collectivity of diffusion we use the isotope effect and for the
dynamic heterogeneity the non-Gaussianity. Having established these we turn to
the van-Hove self correlation function that allows separation of slow and fast 
diffusional
motion. Separating the different contributions to the stress auto-correlations we
find the contributions of slow and fast atoms to the viscosity. The different processes
governing diffusion and viscosity are essentially on different time-scales. Comparing
these is essential in understanding the SER in undercooled liquids.

\subsection{Isotope effect}

At high temperatures and low densities, diffusion in liquids is dominated by binary 
collisions. The kinetic approximation for the mass dependence of the diffusion constant,
$D \propto 1/\sqrt{m}$, holds. When the temperature is lowered or the density is 
increased, effects of collective motion gain importance and the diffusional mass 
changes to an effective mass, $D \propto 1/\sqrt{m_{\rm eff}}$.  
A frequently used measure of
this collectivization is the isotope effect parameter $E$\cite{schoen:58} 
\begin{equation}
 E^\ell_{\alpha\beta} = \frac{D^\ell_\alpha/D^\ell_\beta -1}{\sqrt{m^\ell_\beta/m^\ell_\alpha}  -1} 
\label{isotope_eq}
\end{equation}
where $\ell$ denotes the different components and $\alpha$ and $\beta$ denote different
isotopes. Using radio-tracer isotopes of Co values of $E \approx 0.1$ or less have been
measured in metallic glasses and supercooled metallic melts \cite{faupel:90,ehmler:98}. 
These low values, compared to the typical value of $E\approx 0.7$ for vacancy diffusion 
in crystals, is taken as strong evidence of a collective process. Due to experimental
difficulties no systematic study of the temperature dependence is available.
Using large mass differences, early molecular dynamics simulations for hard disks and 
LJ-systems found again small isotope effects \cite{herman:72,ebbsjo:74,bearman:81,nuevo:95}. Using small mass differences, simulations of monotonic and binary LJ-systems at 
pressure $p=0$
gave a drop from $E\approx 0.3$ at high temperatures to $E \approx 0.05$ approaching the
glass transition.\cite{KS:00,S:01}.

We repeated these calculations for the present system that has  a shorter cutoff and is
under high pressure, $p=5$.
We changed the the mass of small randomly chosen subsets of A- and B-atoms, each
comprising 1.8\% of the species, by $\pm \Delta m$. The average mass was thus kept
constant. Due to the small concentrations clustering effects should not be important.
For all 6 atom species (A and B, average mass, heavy and light) the diffusion coefficient
was calculated and $E^\ell_{\alpha\beta}$ was evaluated. The mass change was $\Delta m =
0.2$. Additional test runs with $\Delta m = 0.1$ and $\Delta m = 0.4$ showed no 
significant difference. Starting point were the 8 samples which had been aged before
the calculation of the diffusion coefficients, Fig.~\ref{diffusion_fig}.
To gain sufficient statistics for each of these 8 samples at least 1000 sets with
changed mass were created and evaluated. A conservative estimate of the resulting
uncertainty is about 10\% and 20\% for the A- and B-atoms,respectively.  

\begin{figure}[htb]
\includegraphics*[bb=40 100 540 549,width=10cm,clip,keepaspectratio]{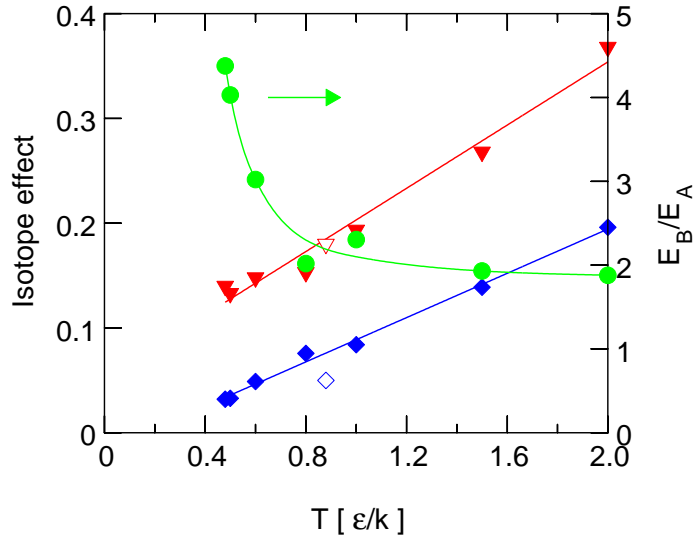}\\[0.5cm]
\caption{Isotope Effect, blue diamonds: A-atoms, red triangles: B-atoms, lines: linear fit. Open symbols: estimates from previous calculation with longer range interaction. Green
circles depict the ratio of the isotope effects of the two components, $E^B/E^A$. The 
line is a guides to the eye.}
\label{isotope_fig}
\end{figure}
Fig.~\ref{isotope_fig} shows a behavior similar to the zero pressure system. The
shorter cutoff reduces the collectivity slightly. In the previous calculations we
have shown that the isotope effect of the majority component A is essentially given by the density
alone. For the minority component B such a scaling with only the density is, however,
not valid. There is both a temperature and density dependence. In general one cannot
expect a pure density scaling for tracer atoms in a multi-component system. 

\begin{figure}[htb]
\includegraphics*[bb=40 100 540 549,width=8cm,clip,keepaspectratio]{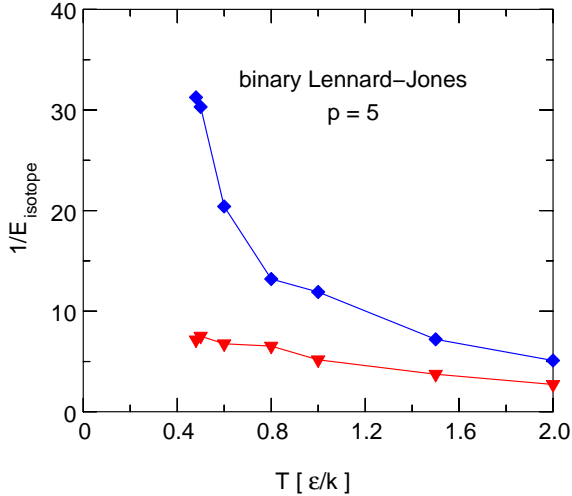}\\[0.5cm]
\caption{Inverse isotope Effect, blue diamonds: A-atoms, red triangles: B-atoms, The lines are a guide to the eyes.}
\label{inverse_isotope_fig}
\end{figure}
The isotope effect can be used to estimate the number of atoms effectively participating 
in the elementary
process of diffusion. We write the effective diffusional mass as 
\begin{equation}
\left(m^\ell_\alpha\right)_{\rm eff} = m^\ell_\alpha +(N^\ell_{\rm D}-1){\overline m}
\label{m_eff_eq}
\end{equation}
where $N^\ell_{\rm D}$ stands for the effective number of atoms moving cooperatively and
${\overline m}$ is the average atomic mass, in our case ${\overline m}=1$.
Inserting the effective mass into Eq.~(\ref{isotope_eq}) one finds $E^\ell \to 1/N_D$.
From Fig.~\ref{isotope_fig} we deduct that there is already considerable collectivity
at the onset temperature $T=1$ and it
increases to more than $N^A_{\rm D}=20$ below $T=0.5$. Converting this into a correlation
length by $N^A_{\rm D}= \rho l_{\rm corr}^3$ we get a correlation length 
$l_{\rm corr}=2.7$ which coincides with the length calculated from the four-point
correlation. If one attributes the isotope effect to string- or chain-like motion
by $N^A_{\rm D}= \rho l_{\rm chain}^{1.6}$ we get a chain length  $l_{\rm chain} \approx 7$.
Here we assumed an effective dimension of $1.5$ for the chain.\cite{OS:99} 

The temperature dependence of $E^\ell$ does not show any 
pronounced feature in Fig.~\ref{isotope_fig}. The change in dynamics is however 
reflected in the inverse isotope effect $1/E^\ell$, Fig.~\ref{inverse_isotope_fig}.
The rapid increase in collectivity in the diffusion of the A-atoms at low temperatures
is evident. Whether the change in dynamics around $T=1$ observed in Fig.~\ref{dadb_fig}
is also reflected in $1/E^\ell$ is beyond our accuracy. The same holds for the ratio 
$E^B/E^A$ in Fig.~\ref{isotope_fig}.
The isotope effect does not
give direct information on the nature of collectivity. Values of $E\approx0.3$ can still
be imagined as originating from weakly correlated pushing in the dense liquid. 
However, that is not likely for values $E^\ell <0.1$ or $N_D >10$.
It is well established that in undercooled densely packed metallic liquids 
collective motion  by chain- (string-)like structures becomes 
dominant.\cite{SOL:93,SGO:97,donati:98} Such mobile chains are the main 
contributor to the dynamic heterogeneity. 



\subsection{Non-Gaussianity}

In glasses and undercooled melts the mobility of the atoms varies in time, the so called
dynamic heterogeneity. Two approaches are commonly used to describe this phenomenon.
Placing the emphasis on the slow particles, four point correlations of displacement or
overlap functions are studied and the dynamic heterogeneity is defined from the
dynamic susceptibility \cite{donati:99,flenner:11}. Alternatively one quantifies
the deviations from Gaussian distributions of displacements, typical for homogeneous
diffusion. We adopt the latter approach due to its direct connections to the 
van-Hove self correlation and to diffusion
where the weight is on the fast particles. 

We define a non-Gaussianity parameter \cite{rahmann:64} 
\begin{equation}
\alpha_2(t) = \frac{3 <s^4(t)>}{5 <s^2(t)>^2} -1. 
\label{nongauss_eq}
\end{equation}
For a purely homogeneous motion $\alpha_2=0$. For heterogeneous motion  $\alpha_2$ 
increases with time. Since undercooled melts are ergodic and in the long time limit 
homogeneous, $\alpha_2 \to 0$ for $t\to 0$.

\begin{figure}[h]
\includegraphics*[bb=40 100 540 549,width=10cm,clip,keepaspectratio]{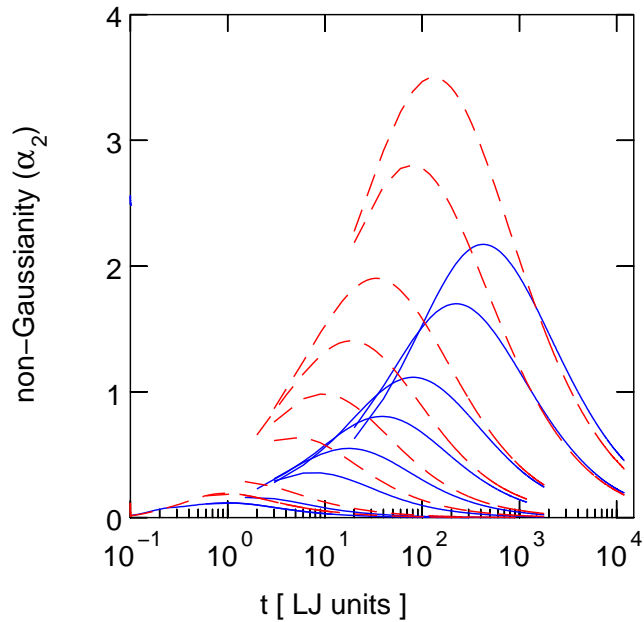}\\[0.5cm]
\caption{Non-Gaussianity parameter; blue solid lines: A-atoms, red dashed lines: B-atoms;
temperatures from top to bottom $T=0.47,0.48.0.50,0.52,0.55,0.60,0.80,1.0$.}
\label{nongauss_fig}
\end{figure}
Fig.~\ref{nongauss_fig} shows the general behavior expected from numerous earlier
simulations. At short times ($t < 1$) there is
a small increase to around 0.1 due to the inhomogeneity of the vibrational spectrum. 
This increase shows little temperature dependence. For longer times the non-Gaussianity 
first increases rapidly, as shown previously\cite{CMS:00}  
$\alpha_2(t) \propto \sqrt{t}$, 
goes through a maximum and finally decays as $\alpha_2(t) \propto 1/t$. The dynamics of 
the smaller B-atoms is much faster than the one of the A-atoms and the heterogeneity is
larger. However, for long times their $\alpha_2(t)$ values decay to the corresponding 
ones of the
A-atoms which shows that the dynamics of the A- and B-particles is coupled. 
We define a non-Gaussianity time as the time when $\alpha_2(t)$ reaches its maximum 
value $\alpha_2^{\rm max} = \alpha_2(t_{\rm NG})$.
For the majority A-particles, by scaling with $t_{\rm NG}$ and $\alpha_2^{\rm max}$, 
the curves of $\alpha_2(t)$ for the
different temperatures can be collapsed to a master curve. 
The scaling does not fully account for the detailed shape
of the maxima. Considering the change in collectivity with temperature, discussed above,
this is not too surprising. The scaling implies that the maximum value of $\alpha_2(t)$
for a given temperature increases $\propto \sqrt{t_{\rm NG}(T)}$ with the time
the maximum is reached. For the A-particles we show $t_{\rm NG}(T)$ in 
Fig.~\ref{times_fig} further down.
For temperatures 
below $T\approx 0.6$, the times $t_{\rm NG}(T)$ increase rapidly and reach at $T=0.47$ values of $t_{\rm NG}=$ 430 and 150, for the A and B particles, respectively. These
times are one order of magnitude less than the diffusion times $t_D$. The dynamic
heterogeneity reaches its maximum long before the particles have on average diffused
over appreciable distances.
We have previously shown \cite{CMS:00}, that the time dependencies of $\alpha_2(t)$ can 
be explained by collective chain- (string-)dynamics. This dynamics involves two times,
one defining the mobility of the chains, the other the decay of the chains. The
times $t_{\rm NG}$ result from an interplay of these two. One can imagine the
heterogeneous dynamics as strings of particles moving in a slow environment. The fast
strings will loose from time to time particles, most likely at the ends, to the slow
environment. To compensate they will pick up new particles. Moving strings also
trigger the formation of new strings or disintegrate. Both scenarios lead to a correlation 
between moving particles\cite{RMP} and cause a transition from slow to fast particle
and vice versa.
 






\subsection{Van-Hove self correlation function} 

\begin{figure}[ht]
\includegraphics*[bb=80 80 550 550,width=10cm,clip,keepaspectratio]{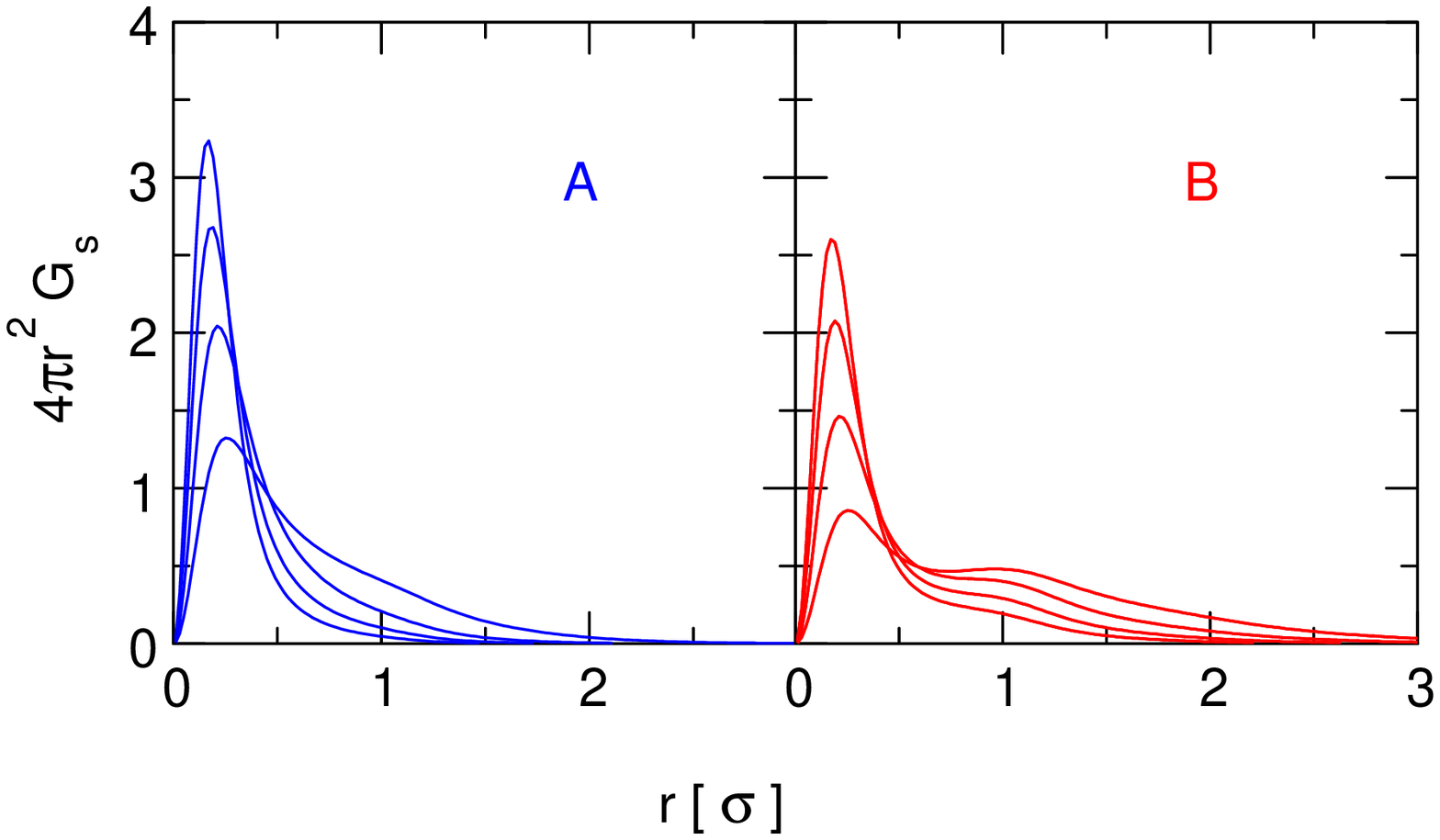}\\[0.5cm]
\vspace{-1.5cm}
\caption{$4\pi r^2$ times the van-Hove self correlation function at $T=0.48$ of A-atoms (left) and B-atoms (right) for times 500, 1000, 2000 and 4000.}
\label{vH_fig}
\end{figure}

The time dependent distribution of the displacements of single atoms can be expressed by the
van-Hove self correlation function (vHSCF). In an isotropic system it can be averaged 
over the space angle to give
\begin{equation}
G^\ell_{\rm s}(r,t) = \frac{1}{4\pi r^2}\frac{1}{N^\ell} \langle \sum_{i=1}^{N^\ell} \delta\left(r - | {\bf R}^i(t)-{\bf R}^i(0) | \right)
\label{vH_eq}
\end{equation}
where ${\bf R}^i(t)$ is the position of atom $i$ at time $t$. $G^\ell_{\rm s}(r,t)$ is a
probability function whose space integral is time independent equal unity. It is usually
plotted as $4 \pi r^2 G^\ell_{\rm s}(r,t)$. For $t=0$ the vHSCF is a $\delta$-function 
at $r=0$. With time the atoms are displaced and the vHSCF broadens. In a purely 
homogeneous system the vHSCF keeps its Gaussian shape, 
the non-Gaussianity $\alpha_2(t)=0$, 
Eq.~(\ref{nongauss_eq}).
Vibrations and
ballistic motion lead to a small broadening which rapidly saturates in time, the 
deviation from a Gaussian shape is small, Fig.~\ref{nongauss_fig}. For long times
the system becomes homogeneous again and the vHSCF is determined by the diffusion
coefficient
\begin{equation}
G^\ell_{\rm s}(r,t) =\left(4\pi D^\ell t\right)^{-3/2} e^{(-r^2/4 D^\ell t)}.
\label{vHdiff_eq}
\end{equation}
At intermediate times, when the non-Gaussianity parameter is large, one observes a
strong deviation from the Gaussian shape. Such deviations are typical for many
disordered systems.\cite{chaudhuri:07}
This is the time region which is of main
interest in the present investigation. As example
Fig.~\ref{vH_fig} shows the the time 
evolution at $T=0.48$ of the vHSCF for both components.
It shows the evolution with time of long range tails of the vHSCF for both
components. At $t=4000$ the often observed secondary peak around $r=\sigma$ becomes 
visible for the B-atoms. For the A-atoms there might be traces of a shoulder. In a
simulation of CuZr it has
been shown that the secondary peak is not due to a preferred jump length, but is due to
increased residence times at previous nearest neighbor sites.\cite{KS:00}. The secondary
peak forms at times comparable to the ``diffusion time'', $t_D$, but long after the
non-Gaussianity has passed its maximum value at $t_{\rm NG}$. For $t\to\infty$ the
long range tails become part of a  strongly broadened Gaussian given by 
Eq.~(\ref{vHdiff_eq}).  
We plot $\log G^\ell_{\rm s}(r,t)$ against $r^2$. In such representation a Gaussian is seen as a straight line.

\begin{figure}[h]
\begin{minipage}[t]{0.495\linewidth}
\includegraphics*[bb=40 40 550 550,height=7.5cm,clip,keepaspectratio]{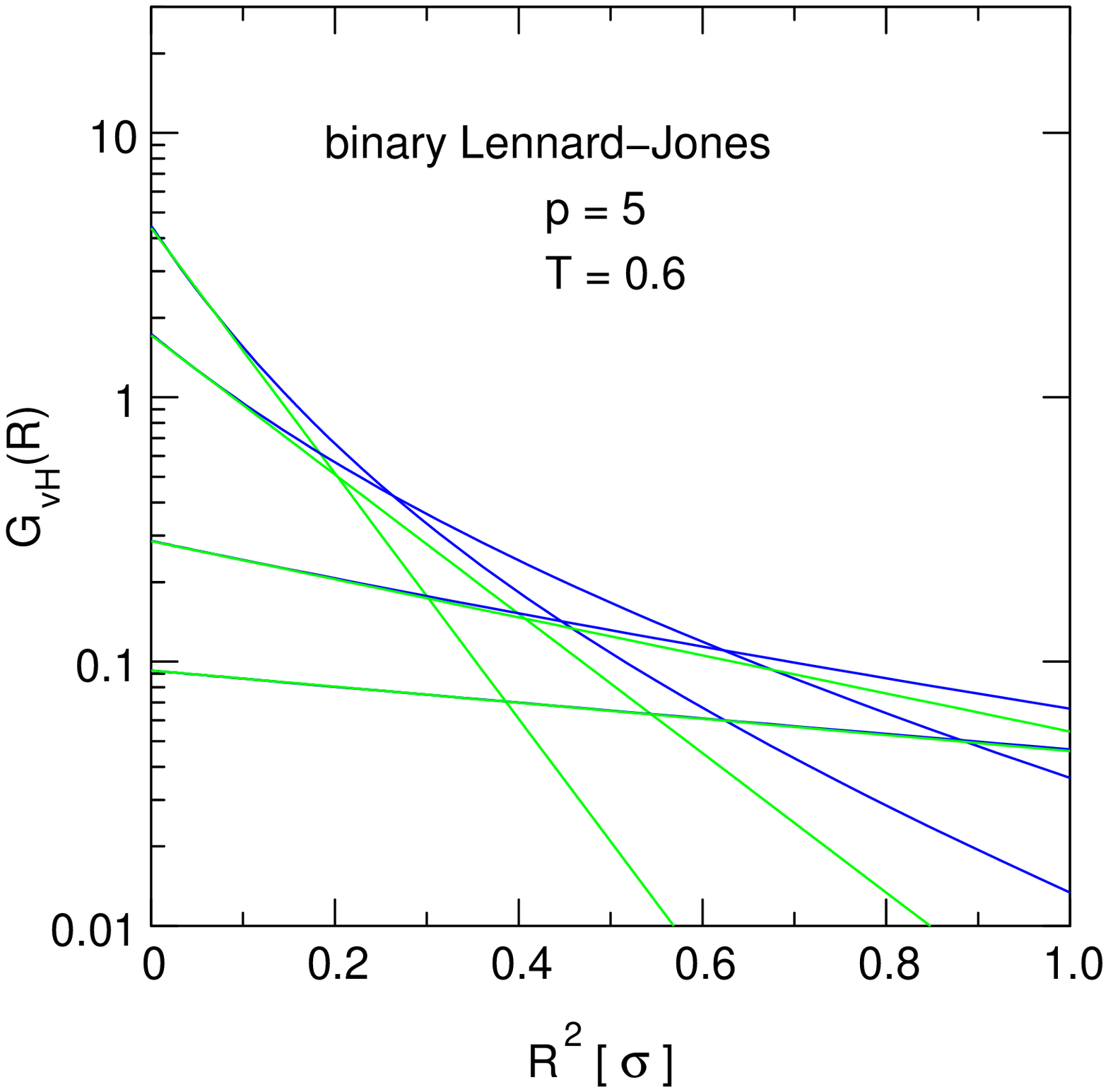}\\[0.5cm]
\end{minipage}\vspace{-1.5cm}
\begin{minipage}[b]{0.495\linewidth}
\includegraphics*[bb=40 40 550 550,height=7.5cm,clip,keepaspectratio]{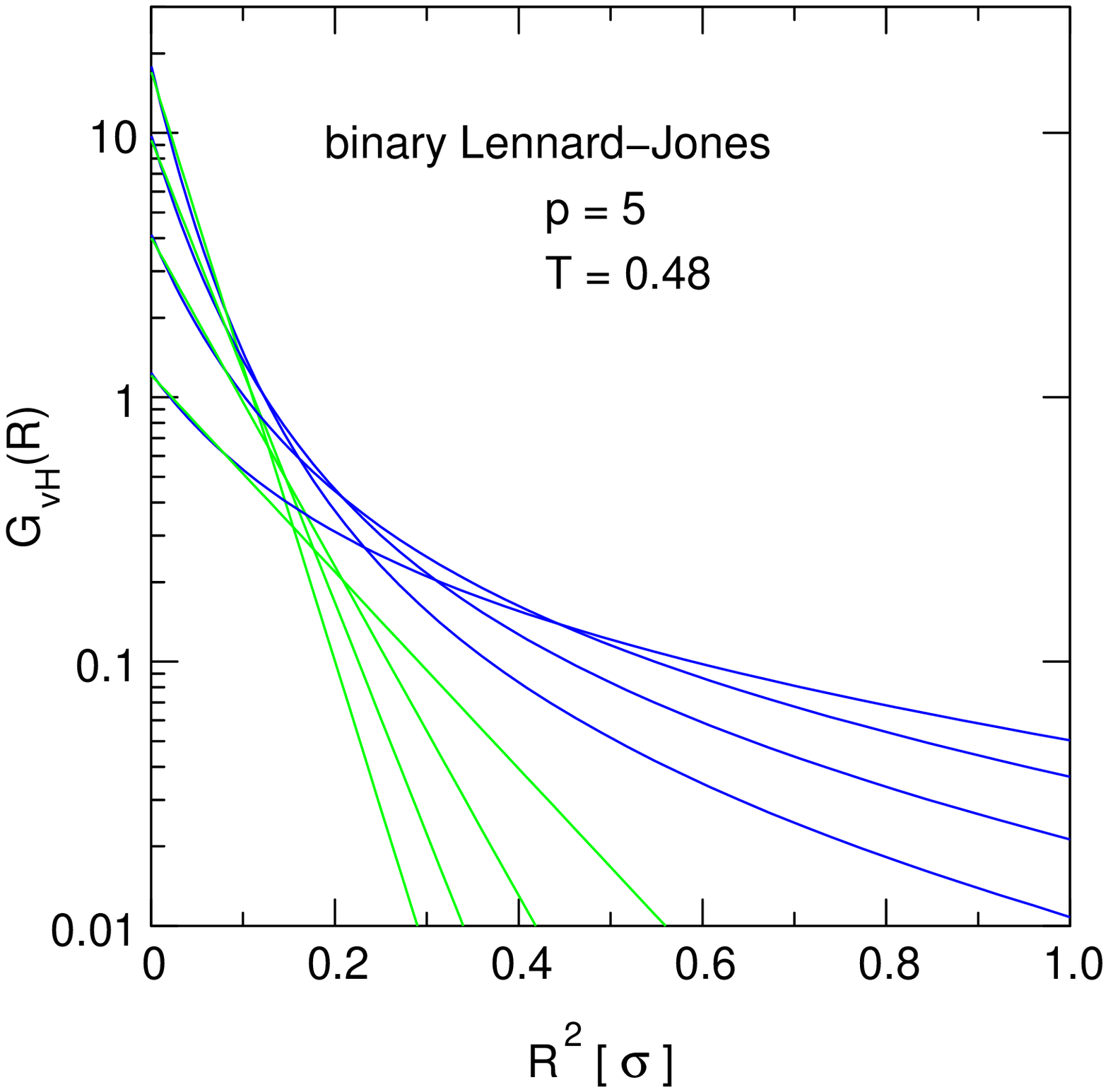}\\[0.5cm]
\end{minipage}
\vspace*{0.5cm}
\caption{Logarithmic plot of van-Hove function of A-atoms against squared distance 
at two temperatures.
Left: $T=0.60$, times $t=$ 15, 30, 90, 120; right:  $T=0.48$ $t=$  500, 1000, 2000 and 4000; calculated values blue, Gaussian fit of central part green straight lines.}
\label{vHlog_fig}
\end{figure}

As example Fig.~\ref{vHlog_fig} shows a logarithmic plot for the A-atoms at $T=0.60$ and$T=0.48$ at
different times. In this representation Gaussians are seen as straight line which
level off with time. The actual values of the vHSCF for larger distances lie above
this straight line, indicating an enhanced mobility of some atoms. The slower than
Gaussian decay of the vHSCF
at larger distances reflects the tails in Fig.~\ref{vH_fig}.
With increasing
time this enhancement vanishes as can be seen for $T=0.60$ at the two later times. The
calculated vHSCF nearly coincides with its Gaussian asymptote, in agreement with the
vanishing non-Gaussianity (Fig.~\ref{nongauss_fig}). However even when there is a
strong curvature in the calculated values,  
the central part still shows a Gaussian shape which persists at the lower temperature 
to 
long times ($t>4000$). 
 The generally accepted picture of the dynamic heterogeneity is that
at any time there are slow and fast atoms which exchange their roles with time and thus
preserve homogeneity in the long time limit. At short times, the central, Gaussian 
part of the vHSCF 
is comprised mainly of those atoms which have no fast history.  
 
We fit this central part of the vHSCF by a Gaussian. 
The fit is done for $R=0.06$ to $R=0.3$. The width of the central
Gaussian is given by the sum of vibration or ballistic motion, cage motion and
slow-diffusion. The first two contributions become constant after some initial
time whereas the diffusional part increases as $4D_{\rm slow}t$. We write  
\begin{equation}
 G_{\rm slow}(r,t) = \frac{A(t)}{(\pi B(t))^{1.5}} \exp{\left[ -r^2/B(t)\right]}.
\label{slowgauss_eq}
\end{equation}
Here $A(t)$ gives the fraction of atoms contained in the central peak. These are 
essentially the atoms which have not jumped (have not been fast atoms).
Such fits were done for temperatures ranging from 0.47 to 2.0.
The slow atom Gaussians  $G_{\rm slow}(r,t)$ are given as green straight lines in  
Fig.~\ref{vHlog_fig}.
The width B(t) can be written as
\begin{equation}
B(t) = {\rm const} + 4D_{\rm slow} t.
\label{width_eq}
\end{equation}
Here the constant accounts for vibrational, ballistic and cage motion which are
supposed to be fast on the diffusional time scales. The above decomposition can only
be done when there is a sufficient time interval during which the slow motion of atoms 
persists 
before these atoms convert into ``fast'' ones. In our case we can evaluate 
Eq.~\ref{width_eq} for temperatures below $T=0.8$. Eq.~\ref{width_eq} can only be
used as long as the majority of atoms is still $slow$. From Fig.~\ref{vHlog_fig} and
the equivalent Figs. for the other temperatures we estimate an accuracy of about 5\%
for the asymptotic $D_{\rm slow}$.
If one does the fit formally to
long times $D_{\rm slow}$ will transform with time into the long time diffusion 
coefficient $D$.

\begin{figure}[h]
\includegraphics*[bb=40 40 550 550,width=10cm,clip,keepaspectratio]{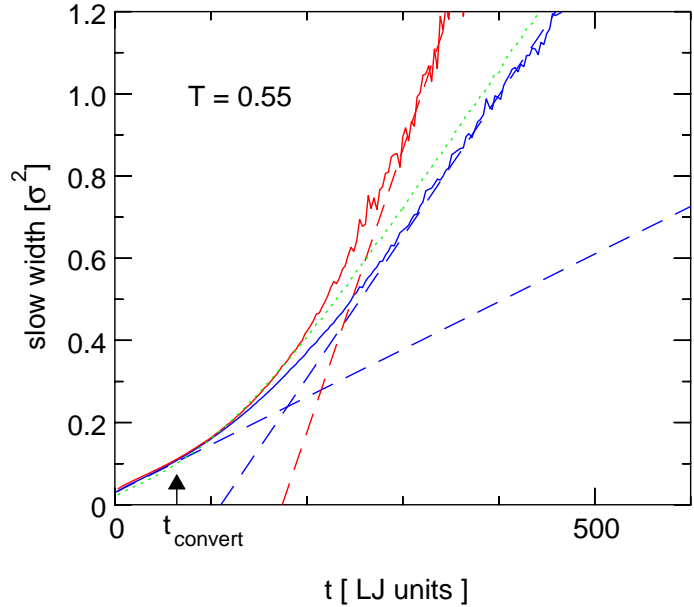}\\[0.5cm]
\vspace{-1.5cm}
\caption{Evolution of the Gaussian width $B(t)$ at $T=0.55$ for A-atoms (blue) 
and B-atoms
(red). The dashed lines indicate the contributions of the slow diffusion at short times 
and the ones calculated from the long time diffusion. The green dotted lines is the width
of the van-Hove function when it is decayed to $1/e$. The decay time $t_{\rm decay}$ 
calculated
from Fig.~\ref{amplitude_fig} is indicated by an arrow.} 
\label{widthoft_fig}
\end{figure}

This time evolution of the Gaussian width of the vHSCF, as calculated from its central 
part, is exemplified in Fig.~\ref{widthoft_fig}.
It shows for both components the changeover from ``slow diffusion'' to normal
diffusion. Neglecting dynamic heterogeneity and doing Gaussian fits for the vHSCF
on some not too long timescale
can give apparent diffusion coefficients varying between
the slow value and the long time diffusion coefficient. Thus values of the diffusivity,
which are derived by a Gaussian approximation, depend on
measuring time, fit range and weighting factors.

At short times the width for A and B particles coincide within our accuracy. This
indicates a strong cooperativity in this temperature range, in agreement with the
above results of non-Gaussianity and isotope effect.

The slow diffusion coefficients derived from Eq.~(\ref{width_eq}) are shown in 
Fig.\ref{diffusion_fig} by green circles. The drop of $D_{\rm slow}$
from the Arrhenius values 
is much more 
pronounced than the one of the average long time coefficients. At $T=0.48$ i
$D_{\rm slow}$ is one order
of magnitude smaller than $D_A$. The product $D_{\rm slow}(T)\eta(T)/T$ strongly drops
below the constant value predicted by the SER, shown by green circles in 
Fig.~\ref{SER_fig}. 
We cannot verify the claim, sometimes made, that
the SER holds for the slow particles. To restore the validity of the SER one could 
introduce an appropriate definition
of ``slow'' by either prescribing a timescale for the fit of the vHSCF or by using 
appropriate cutoff radii. We take a different approach and consider the heterogeneity
of $\eta(T)$ as well as of $D(T)$.

\begin{figure}[htb]
\includegraphics*[bb=40 40 550 550,width=8cm,clip,keepaspectratio]{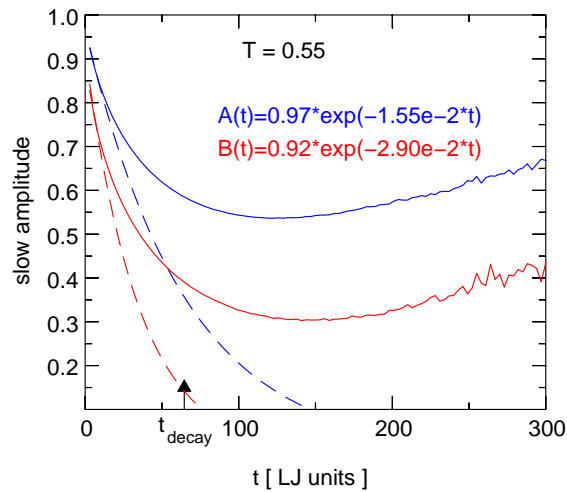}\\[0.5cm]
\vspace{-1.5cm}
\caption{Evolution of the amplitude of the Gaussian width $A(t)$ at $T=0.55$ for 
A-atoms (blue) and B-atoms (red); full lines calculated values, dashed lines exponential
fit to short time decay.} 
\label{amplitude_fig}
\end{figure}

As mentioned before an essential parameter characterizing the heterogeneous dynamics
is the conversion rate from slow to fast.
The decay of the amplitude of the central Gaussian can be used to extract this rate.
For short times the decay is given by
\begin{equation}
A(t) = A_0 \exp{\left[-t/t_{\rm convert}\right]}.
\label{amplitude_eq}
\end{equation}
A value $A_0=1$ indicates that all atoms are in the central Gaussian. Fast cage hopping 
and heterogeneity of vibration and ballistic motion reduce the factor. As
the central Gaussian merges with time into the long time diffusional Gaussian so does
$A(t)$ after the initial decay increase again to one. 
The amplitude $A(t)$ initially decays exponentially indicating the rate of the
transformation of slow particles into fast ones. Assuming that this is correlated
to jumps, chains of atoms losing atoms and picking up others, this gives an estimate
of the jump rate. An example of the time dependence of the central Gaussian is shown in
Fig.~\ref{amplitude_fig}. The conversion time $t_{\rm convert}$ increases in the 
temperature interval $1.7 < 1/T < 2.1$ by two orders of magnitude, see 
Fig.~\ref{times_fig}. It is much larger than the non-Gaussianity time but smaller than
the diffusion time. Fig.~\ref{widthoft_fig} shows that the central Gaussian showing the
slow diffusion is clearly visible for time up to and even above $t_{\rm convert}$. 

\subsection{Viscosity time and partial viscosities}

The Green-Kubo expression for the viscosity offers two advantages. First, one can study
the time evolution of the Green-Kubo integral Eq.~(\ref{eta_eq})
\begin{equation}
\eta(T,t) = \int_0^t \hat{\eta}(T,t') dt'
\label{etatime_eq}
\end{equation}
and thus define a time
during which the stresses are sufficiently correlated to contribute to $\eta(T)$. To
quantify the time-span during which the major part of $\eta$ is accumulated we
define a viscosity time as
\begin{equation}
\int_0^{t_{\rm visc}} \hat{\eta}(T,t) dt = \eta(T,t_{\rm visc})= 0.8 \;\eta(T)
\label{t_visc_eq}
\end{equation}
Secondly the sums in Eq.~(\ref{eta_eq}) can be split to show the contributions of
different groups of atoms to $\eta(T)$.

\begin{figure}[h]
\includegraphics*[bb=00 40 550 550,width=10cm,clip,keepaspectratio]{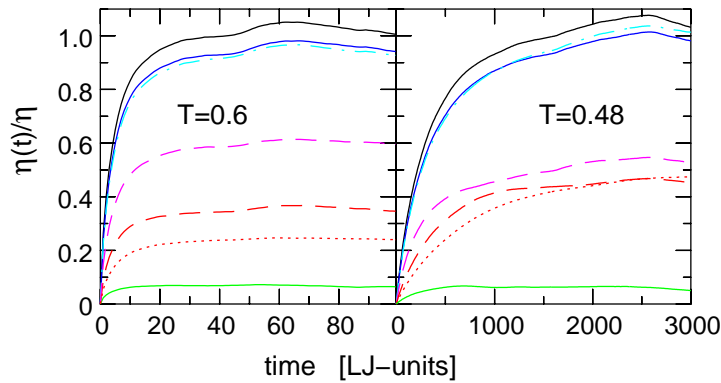}\\[0.5cm]
\vspace{-1.5cm}
\caption{Time evolution of $\int^t \hat{\eta}(T,t')dt'$ and 
 $\int^t \tilde{\eta}_{g1,g2}(T,t')dt'$ normalized by $\eta(T)$, 
for two temperatures. From top to bottom:
solid lines: total-total (black), A-total (blue), B-total (green); 
dash-dotted: A-A (cyan), dashed lines: fast-total (magenta), slow-total (red); dotted
line: slow-slow.} 
\label{eta_t_fig}
\end{figure}

The time evolution of $\eta(T,t)/\eta$ is shown for two examples 
in Fig.~\ref{eta_t_fig} (solid black line). From these time evolutions we gain
$t_{\rm visc}(T)$ shown as red up-triangles in Fig.~\ref{times_fig}. It shows that
the stress correlation decays faster,
the viscosity evolves faster, than the conversion of a slow particle to a fast one,
$t_{\rm visc}<t_{\rm convert}$, whence the stress 
evolution is
heterogeneous. Slow environments remain slow over relevant times. It is therefore
useful to study the different contributions to $\eta(T)$ separately. We introduce
partial stress tensors for groups of atoms
\begin{equation}
\tilde{\sigma}^{xy}_g = \sum_{i=1}^{N_g} \left( m_i v_i^x v_i^y - 
                          \sum_{j>i}{
                         \frac{\partial V_{ij}}{\partial r_{ij}} 
                         \frac{r_{ij}^x r_{ij}^y} {r_{ij}}}
\right).
\label{partial_stress_eq}
\end{equation}
and 
\begin{equation}
\tilde{\eta}_{g1,g2}(T,t) = \frac{1}{k_b T V}\langle \tilde{\sigma}^{xy}_{g1}(t)\tilde{\sigma}^{xy}_{g2}(0)\rangle ,
\label{partial_etat_eq}
\end{equation}

where the index $g$ denotes all particles (total), all A or B particles (A or B), all
slow A-particles or all fast A-particles. By definition $\tilde{\eta}_{\rm total,total}(T,t) = \hat{\eta}(T,t)$.
We count those $N_{\rm slow} = N_A / e$ 
A-atoms as slow which have been displaced least during the time $t_{\rm convert}$. 
As we can determine $t_{\rm convert}$ only for temperatures $T\le 0.6$, we only
analyze the slow and fast contributions in that temperature range.

As example Fig~\ref{eta_t_fig} shows the normalized $\int \hat{\eta}(T,t')dt'$ and some 
of the constituting terms $\int \tilde{\eta}_{g1,g2}(T,t')dt'$
for two temperatures.  
Not surprisingly we find that the viscosity of our system is dominated at all 
temperatures by the 80\% A-particles. The contribution of the B-particles (B-total) 
is halved from
about 12\% at $T=2$ to $6\%$ at $T=0.48$. 
We will, therefore, concentrate on the A-particles. Comparing the A-total and A-A terms 
one sees that the A-B contribution is nearly negligible. 
Since we have a ratio of 1.7 between the numbers of fast and slow A-atoms one expects
in a homogeneous system a similar ratio for the fast-total and slow-total contributions
to $\int \eta(T,t') dt'$. This holds approximately at $T=0.6$. But at the lower 
temperature
$T=0.48$ the two contributions become comparable. Furthermore the slow-slow contribution
becomes equal to the slow-total one, i.e. the slow A-atoms act as a subsystem. 
We use this 
to introduce a virtual slow system where all A-atoms are slow and contribute, as the
slow atoms contribute in the real system. We substitute 
A-A $\to$ slow-slow $\times$ $(N_A/N_{\rm slow})^2$. The resulting viscosity of the
virtual slow system is shown in Fig.~\ref{viscosity_fig} by green circles. It obeys 
approximately the SER. In this virtual system the stress correlations decays more slowly
than in the real system, the virtual $t_{\rm visc}$ becomes similar to $t_{\rm convert}$.

\subsection{Timescales}

\begin{figure}[htb]
\includegraphics*[bb=40 40 550 550,width=8cm,clip,keepaspectratio]{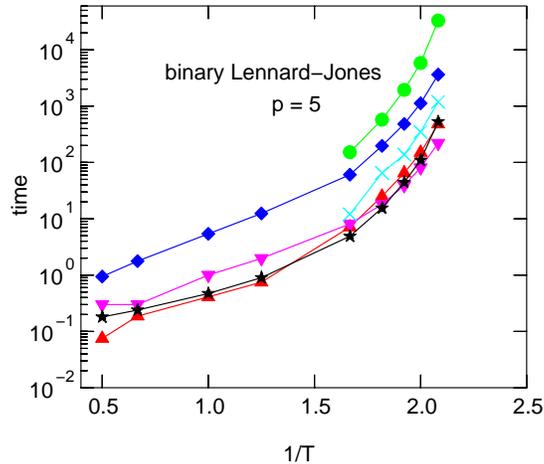}\\[0.5cm]
\vspace{-1.5cm}
\caption{Temperature dependence of the different time scales: 
$t_\alpha$ (black asterisks),
non-Gaussianity maximum, $t_{\rm NG}$, (magenta down triangles),
diffusion time of A-atoms, $t_{\rm D}$, (blue diamonds),
slow diffusion time,$t_{\rm D_{slow}}$, (green circles),
viscosity time, $t_{\rm visc}$, (red up triangles),
decay time of slow atom concentration, $t_{\rm convert}$ (cyan crosses).}
\label{times_fig}
\end{figure}

In Fig.~\ref{times_fig} we summarize the different timescales encountered in this
investigation and compare them with the $\alpha$-relaxation time $t_\alpha$. 
All timescales show the characteristic upturn at temperatures  below $T\approx 0.6$. 
The time $t_{\rm visc}$, which measures essentially the lifetime of the stress correlation
entering the Green-Kubo relation for the viscosity, broadly coincides with $t_\alpha$.
The maximum of the non-Gaussianity is reached on similar time scales. For the highest
temperatures $t_{\rm NG}$ saturates when it is no longer dominated by diffusion and
the non-Gaussianity is given by vibrational (ballistic) heterogeneity. At the low
temperature side $t_{\rm NG}$ drops below $t_\alpha$ as has been noted earlier 
\cite{saltzman:06}. With decreasing temperature the slow-diffusion time, 
$t_{D_{\rm slow}}$,
markedly diverges from the average diffusion time, $t_{\rm D}$, and from $t_\alpha$.
The most interesting result is that the timescale for conversion from slow to fast 
A-particles, $t_{\rm convert}$ is larger than
$t_\alpha$ and the average  $t_{\rm visc}$. Dividing the atoms into slow and fast
is therefore sensible on the timescales relevant for the buildup of the viscosity.
There are of course some stress correlations over longer times, but these contribute
only little to the viscosity.  
We want to stress that $t_{\rm convert}$ is not the same as the time a particle is a
fast diffuser $t_{\rm fast}$, as studied before\cite{yamamotu:98}. The two times are
related by an equilibrium condition for the concentration of slow and fast particles,
$c_{\rm slow}/t_{\rm convert}=c_{\rm_fast}/t_{rm fast}$. 
In  the virtual system that consists only
of slow A-particles $t_{\rm visc}$ is increased  to approximately  $t_{\rm convert}$ of
the real system. 

\subsection{Error-analysis}

The accuracy of our results are affected by both limitations of the computations and
by system immanent problems. The statistical errors can be estimated from the 
scatter of the data points in time as well as temperature. Due to the long aging
times the samples at the different temperatures can be taken as fairly independent.
We estimate the statistical error of the viscosity as less than 5\% and for the partial
viscosities as less than 10\%. The diffusivities are calculated more accurately 
from the mean square displacements. At the lowest temperatures aging effects become
noticeable. 

It has been shown that at low temperatures there is a split between the
aging rates of diffusivity and heterogeneity on one side and energy and pressure or
volume on the other, with the second being the slower ones.\cite{S:12} Viscosity
presumably belongs to the slower aging group. Insufficient aging then might result
in good values for the diffusivity but too low ones for the viscosity and as effect
too low values of the SER. For our data this might have occurred for $T=0.47$. 
Therefore we did not use this temperature in the final analysis but restricted to
temperatures up from $T=0.48$.       

Both the calculated diffusion coefficients and the viscosities are affected by 
finite size effects which increase with lowering the temperature. These size
effects are more pronounced for the diffusivities \cite{yeh:04}. Simulation
with different system sizes have shown that for our systems with 5488 particles
the effect on the calculated SER is only marginal.

More important are the system immanent uncertainties of our calculation. We have
introduced some times which define timescales but do not affect the actual 
calculations ($t_D$, $t_{D_{\rm slow}}$, $t_{\rm visc}$, $t_{\rm NG}$). Inaccuracies
in these numbers do not affect the general results. The conversation time from
slow to fast ($t_{\rm convert}$) is more critical. It is determined from the
$r=0$ values of the low $r$ asymptotes in Fig.~\ref{vHlog_fig}. At the higher
temperature ($t>0/8$) the time span between ballistic or vibrational motion and
significant long range diffusion is too short for an accurate evaluation and we
have omitted these values. According to Eq.~(\ref{vHdiff_eq}) an error in the 
$D_{\rm slow}$ propagates with a factor 1.5 to $t_{\rm convert}$. Apart from the lowest 
value we don't expect the error to exceed a few percent.

The most critical step is the identification of the slow particles. During the
relevant time for the viscosity $t_{\rm viscous}$ a large number of particles 
are for some time ``fast''. These particles also contribute to $G_s^\ell(r=0,t)$,
Fig.~\ref{amplitude_fig}, but should not be counted as ``slow''. We have taken
this into account by picking out these atoms which have moved least, including
reversed jumps. Varying, in reasonable limits, $t_{\rm convert}$ and its fraction used 
to determine $N_{\rm slow}$ the SER for the slow system is changed by around 20\%.  

\section{Conclusion}
Using molecular dynamics simulation of a binary Lennard-Jones melt
we have investigated diffusion, viscosity and the Stokes-Einstein relation (SER),
between them, as function of temperature. Three temperature regimes can be distinguished.
At high temperatures the diffusion obeys an Arrhenius law, the viscosity is low and the
SER holds ($D\eta/T= {\rm const}$). Upon cooling the dynamics becomes
increasingly collective but not yet strongly heterogeneous. In this intermediate 
temperature the SER is weakly violated. The relation between viscosity and diffusivity
is often described by a fractional SER, $D\eta/T^p = {\rm const}$.\cite{pollack:85}
For temperatures down to $T=0.6$ our results can be fitted with $p=0.8$, in good 
agreement with experiments on organic liquids\cite{fujara:92,swallen:03}. As discussed
in the introduction,
fractional SER have been derived for numerous models accounting for hopping and
fluctuations. As alternative a transition from the SER to a relation 
$D\eta = {\rm const}$, emphasizing the growing collectivity, has been 
claimed.\cite{brillo:11,HS:11,HS:16} The present data do not suffice to decide
between the two description. The emphasis of this work is on the lower temperature
region where the SER violation is much stronger and cannot be described by a 
fractional SER with a positive exponent $p$.

At low temperatures approaching the glass transition the viscosity increases
rapidly and the diffusion drops below the Arrhenius values, 
the SER is strongly violated,  $D\eta/T$ rapidly increases with $1/T$. At
$T=0.48$ the SER is violated by a factor of 2.5. The system shows both large 
dynamic heterogeneities and a strong collectivity of motion. We show that the
strong violation of the SER results from the heterogeneity. With decreasing
temperature viscosity becomes dominated by slow particles while diffusion is
by fast particles. We share this picture with the majority of workers in the
field. The definition of fast and slow often is limited to hopping motion. Other
than in hard sphere systems, in metallic melts there is a smooth distribution 
of hopping distances and, therefore there is no clear identification of particles 
which have hopped.\cite{KS:04} As remedy sometimes a cutoff length in the
van-Hove self-correlation (vHSCF) is used to separate slow and fast particles. 
The results for the SER depend crucially on this cutoff.

In the present work we show that slow and fast particles can be separated using 
the vHSCF. No distinction between hopping and flow motion is needed.
At not too large times the vHSCF exhibits a Gaussian center. This is due to
slow particles whereas fast ones are mainly seen in the non-Gaussian tails. 
The broadening
of the central Gaussian with time measures a slow diffusivity. From the amplitude
of the Gaussian we gain the lifetime of a particle as
slow, before it undergoes a fast motion which takes it out of central Gaussian.
This conversion time is longer than timespan during which the integral over the
stress correlation contributes strongly  in the
Green Kubo formula
to the viscosity. On the relevant time scales we observe a slow subsystem that
with decreasing temperature more and more dominates the viscosity. It acts as a
quasi static heterogeneity. Calculating an SER from the diffusivity of the slow
particles and the viscosity leads to a dramatic underestimation of the SER values.
Making a Gaussian approximation of the vHSCF over some distance would increase the
SER ratio again. The SER can be regained without any implicit parameter when
one considers the heterogeneity of both he stress correlation
as of the diffusivity. In a virtual melt formed from the slow 
subsystem the SER is obeyed.

\section{Acknowledgments}
We are grateful to T. Voigtmann for many stimulating discussions throughout this work. 
HRS acknowledges the hospitality
of the Institut f\"ur Materialphysik im Weltall at DLR, Cologne. The authors gratefully 
acknowledge the computing time granted on the supercomputer JUROPA at J\"ulich 
Supercomputing Center (JSC). The work was partially supported by the German Academic
Exchange Servive (DAAD) through the DLR-DAAD programme under grant No. 131.


\begin{thebibliography}{99}
\bibitem{einstein} A. Einstein, Ann. Phys. {\bf 17}, 549 (1905).
\bibitem{balucani} U. Balucani and M. Zoppi, {\it Dynamics of the Liquid State}
(Clarendon Press, Oxford, 1994).
\bibitem{clough:82} G. Clough, J. Physiol. {\bf 328}, 389 (1982).
\bibitem{mccarthy:01} M.R. McCarthy, K.D. Vandegriff, and R. M. Winslow, Biophys. Chem.
{\bf 94}, 103 (2001).
\bibitem{poirier:88} J. P. Poirier, Geophys. J. {\bf 92}, 99 (1988).
\bibitem {poe:97} B. T. Poe, P. F. McMillan, D. C. Rubie, S. Chakraborty, J. Yarger,
and J. Diefenbacher, Science {\bf 276}, 1245 (1997). 
\bibitem{perrin:34} F. Perrin,  J. Phys. Radium {\bf 5}, 497 (1934). 
\bibitem{gaskell:89} T. Gaskell, U. Balucani,and R. Vallauri, Phys. Chem. Liq. {\bf 19},
193 (1989).

\bibitem{shi:13} Z. Shi, P. G. Debenedetti, and F. Stillinger, J. Chem. Phys. {\bf 138},
12A526 (2013).
\bibitem{roessler:90} E. R\"ossler, Phys. Rev. Lett. {\bf 65}, 1595 (1990).
\bibitem{fujara:92} F. Fujara, B. Geil, H. Sillescu, and G. Fleischer, Z. Phys. B 
{\bf 88}, 195 (1992).
\bibitem{cicerone:95} M. T. Cicerone, F. R. Blackburn, and M. D. Ediger, J. Chem. Phys.
{\bf 102}, 471 (1995).
\bibitem{heuberger:96} G. Heuberger and H. Sillescu, J. Phys. Chem. {\bf 100}, 
15255 (1996).
\bibitem{voronel:98} A. Voronel, E. Veliyulin, V. Sh. Machavariani, A. Kisliuk, and 
D. Quitmann, Phys. Rev. Lett. {\bf 80}, 2630 (1998).
\bibitem{meyer:03} A. Meyer, W. Petry, M. Koza, and M.-P. Macht, Appl. Phys. Lett.
{\bf 83}, 3894 (2003).
\bibitem{swallen:03} S. F. Swallen, P. A. Bonvallet, R. J. McMahon, and M. D. Ediger,
Phys. Rev. Lett. {\bf 90}, 015901 (2003).
\bibitem{bartsch:10} A. Bartsch, K. R\"atzke, A. Meyer, and F. Faupel, Phys.Rev. Lett.
{\bf 104}, 195901 (2010). 
\bibitem{brillo:11} J. Brillo, A. I. Pommrich, and A. Meyer, Phys. Rev. Lett. {\bf 107},
165902 (2011).
\bibitem{thirumalai:93} D. Thirumalai and R. D. Mountain, Phys. Rev. E {\bf 47},
479 (1993).
\bibitem{nicodemi:98} M. Nicodemi and A. Coniglio, Phys. Rev. E {\bf 57}, R39 (1999).
\bibitem{angelani:98} L. Angelani, G. Parisi, G. Ruocco, and G. Viliani, Phys. Rev.
Lett. {\bf 81}, 4648 (1998).
\bibitem{yamamotu:98} R. Yamamotu and A. Onuki, Phys. Rev. E {\bf 58}, 3515 (1999).
\bibitem{allegrini:99} P. Allegrini, J. F. Douglas, and S. C. Glotzer, Phys. Rev. E 
{\bf 60}, 5714(1999).
\bibitem{demichele:01} C. DeMichele and D. Leporini, Phys. Rev. E {\bf 63}, 
036701 (2001).
\bibitem{mukherjee:02} A. Mukherjee, S. Bhattacharyya, and B. Bagchi, J. Phys. Chem.
{\bf 116}, 4577 (2002).
\bibitem{bordat:03} P. Bordat, F. Affouard, M. Descamps, and F. M\"uller-Plathe, 
J. Phys.:Condens Matter {\bf 15}, 5397 (2003).
\bibitem{kumar:06} S. K. Kumar, G. Szamel and J. F. Douglas, J. Chem. Phys. {\bf 124},
214501 (2006).
\bibitem{das:08} S. K. Das, J. Horbach, and T. Voigtmann, Phys. Rev. B {\bf 78}, 064208
(2008).
\bibitem{affouard:09} F. Affouard, M. Descamps, L.-C. Valdes, J. Habasaki, P. Bordat,
and K. L. Ngai, J. Chem. Phys. {\bf 131} 104510 (2009).
\bibitem{HS:11} X. J. Han and H. R. Schober, Phys. Rev. B {\bf 83}, 224201 (2011).
\bibitem{HS:16} X. J. Han and H. R. Schober, J. Chem. Phys. {\bf 144}, 124505 (2016).
\bibitem{lu:12} Y. L\"u, H. Cheng, and M. Chen, J. Chem. Phys. {\bf 136}, 214505 (2012).
{\bf 138}, 12A548 (2013).
\bibitem{sengupta:13} S. Sengupta, S. Karmakar, C. Dasgupta, and S. Sastry, J. Chem. Phys.
\bibitem{sengupta:14} S. Sengupta and S. Karmakar, J. Chem. Phys. {\bf 140}, 224505 
(2014).\
\bibitem{jaiswal:15} A. Jaiswal, T. Egami, and Y. Zhang, Phys. Rev. B {\bf 91}, 134204 (2015).
\bibitem{goetze:92} W. Goetze and L. Sjogren, Rep. Prog. Phys. {\bf 55}, 241 (1992).
\bibitem{pollack:85} G. L. Pollack and J. J. Enyeart, Phys. Rev. A {\bf 31}, 980 (1985).
\bibitem{zwanzig:85} R. Zwanzig and K. Harrison, J. Phys. Chem {\bf 83}, 5861 (1985).
\bibitem{heuer:08} A. Heuer, J. Phys.: Condens. Matter {\bf 20}, 373101 (2008).
\bibitem{schweitzer:04} K. S. Schweitzer and E. Saltzman, J. Phys. Chem. B {\bf 108},
19729 (2004).
\bibitem{jung:04} Y. J. Jung, J. P. Garrahan, and D. Chandler, Phys. Rev. E {\bf 69},
061205 (2004).
\bibitem{ngai:99} K. L. Ngai Phil. Mag. B {\bf 79}, 1783 (1999).
\bibitem{rizzo:15} T. Rizzo and T. Voigtmann, EPL {\bf 111}, 56008 (2015).
\bibitem{RMP} F. Faupel, W. Frank, M.-P. Macht, H. Mehrer, V. Naundorf, K. R\"atzke,
H. R. Schober, S. K. Sharma, and H. Teichler, Rev. Mod. Phys. {\bf 75},237 (2003).
\bibitem{KS:04} M. Kluge and H. R. Schober, Phys. Rev. B {\bf 70}, 224209 (2004).
\bibitem{kob:95} W. Kob and H. C. Andersen, Phys. Rev. E {\bf 51}, 4626 (1995).
(Butterworts, London, 1984).
\bibitem{allen:87} M. P. Allen and D. J. Tildesley, {\it Computer Simulation of Liquids},
(Clarendon Press, Oxford 1987).
\bibitem{lammps} http://lammps.sandia.gov.

\bibitem{schoen:58} A. H. Schoen, Phys. Rev. Lett. {\bf 1}, 138 (1958).
\bibitem{faupel:90} F. Faupel, P. W. H\"uppe, and K. R\"atzke, Phys. Rev. Lett. {\bf 65},
1219 (1990).
\bibitem{ehmler:98} H. Ehmler, A. Heesemann, K. R\"atzke, F. Faupel, and U. Geyer, Phys. Rev. Lett.
{\bf 80}, 4919 (1998).
\bibitem{herman:72} P. T. Herman and B. J. Alder, J. Chem. Phys. {\bf 56}, 987 (1972).
\bibitem{ebbsjo:74} I. Ebbsj\"o, P. Schofield, K. Sk\"old, and I. Waller, 
J. Phys. C {\bf 7}, 3891 (1974).
\bibitem{bearman:81} R. J. Bearman and D. L. Jolly, Mol. Phys. {\bf 44}, 665 (1981).
\bibitem{nuevo:95} M. J. Nuevo, J. J. Morales, D. M. Heyes, Phys. Rev. E {\bf51},
2026 (1995). 
\bibitem{KS:00} M. Kluge and H. R. Schober, Phys. Rev. E {\bf 62}, 597 (2000).
\bibitem{S:01} H. R. Schober, Solid state Commun. {\bf 119}, 73 (2001). 
\bibitem{OS:99} C. Oligschleger and H. R. Laird, Phys. Rev. B {\bf 59}, 811 (1999).
\bibitem{SOL:93} H. R. Schober, C. Oligschleger, and B. B. Laird, J. Non-Cryst. Solids
{\bf 156}, 965 (1993).
\bibitem{SGO:97} H. R. Schober, C. Gaukel, and C. Oligschleger, Prog. Theor. Phys. Suppl.
{\bf 126}, 67 (1997).
\bibitem{donati:98} C. Donati, J. F. Douglas, W. Kob, S. J. Plimplton, P. H. Poole, and
S. C. Glotzer, Phys. Rev. Lett. {\bf 80}, 2338 (1998).
\bibitem{donati:99} C. Donati, S. C. Glotzer, and P. H. Poole, Phys. Rev. Lett. {\bf 82},
5064 (1999).
\bibitem{flenner:11} E. Flenner, M. Zhang, and G. Szamel, Phys. Rev. E {\bf 83}, 
051501 (2011).
\bibitem{rahmann:64} A. Rahmann, Phys. Rev. {\bf 136}, A405 (1964).
\bibitem{CMS:00} D. Caprion, J. Matsui, and H. R. Schober, Phys. Rev. Lett. {\bf 85},
4239 (2000).
\bibitem{chaudhuri:07} P. Chaudhuri, L. Berthier, and W. Kob, Phys. Rev. Lett {\bf 99},
060604 (2007).
\bibitem{saltzman:06} E. J. Saltzman and K. S. Schweitzer, Phys. Rev. E {\bf 74},
061501 (2006).
\bibitem{S:12} H. R. Schober, Phys. Rev. B {\bf 85}, 024204 (2012).
\bibitem{yeh:04} I-C. Yeh and G. Hummer, J. Phys. Chem. B {\bf 108}, 15873 (2004).
\end{thebibliography}
\end{document}